\documentclass[journal=ancac3,manuscript=article,layout=twocolumn]{achemso}

\setkeys{acs}{maxauthors = 0}      

\setkeys{acs}{articletitle = true}

\usepackage{comment}
\usepackage{todonotes}
\usepackage{amssymb}
\usepackage{amsfonts}
\usepackage{amsmath}

\usepackage{graphicx}
\usepackage{dcolumn}
\usepackage{bm}

\usepackage[export]{adjustbox}

\usepackage{array}
\newcommand\Tstrut{\rule{0pt}{2.8ex}}
\newcommand\Bstrut{\rule[-1.ex]{0pt}{0pt}}   
\newcolumntype{P}[1]{>{\centering\arraybackslash}p{#1}}

\usepackage[version=3]{mhchem}

\usepackage{graphicx}
\usepackage{dcolumn}
\usepackage{bm}


\newcommand{\df}{\mathrm{d}}

\newcommand{\kt}{k_\text{B}T}
\newcommand{\eps}{\varepsilon}
\newcommand{\lB}{\ell_\text{B}}
\newcommand{\debye}{\lambda_\text{D}}
\newcommand{\lGC}{\ell_\text{GC}}
\newcommand{\rhoe}{\rho_\text{e}}

\newcommand{\phis}{\phi_\text{s}}

\newcommand{\ns}{n_\text{s}}

\newcommand{\sgn}{\text{sgn}}

\newcommand{\uzh}{Department of Chemistry, Universit\"at Z\"urich, 8057 Z\"urich, Switzerland}
\newcommand{\ilm}{Univ Lyon, Univ Claude Bernard Lyon 1, CNRS, Institut Lumi\`ere Mati\`ere, F-69622, VILLEURBANNE, France}
\newcommand{\iuf}{Institut Universitaire de France (IUF), 1 rue Descartes, 75005 Paris, France}
\newcommand{\tuhh}{Hamburg University of Technology, Insitute of Polymers and Composites, Hamburg, 21073, Hamburg}
\newcommand{\helm}{Helmholtz-Zentrum Hereon, Institute of Surface Science, Geesthacht, 21502, Germany}

\author{Laurent Joly}
\affiliation[Universit\'e Lyon 1]{\ilm}
\alsoaffiliation[IUF]{\iuf}
\author{Robert H. Mei{\ss}ner}
\affiliation[Hamburg University of Technology]{\tuhh}
\alsoaffiliation[Helmholtz-Zentrum]{\helm}
\author{Marcella Iannuzzi}
\affiliation[Universit\"at Z\"urich]{\uzh}
\author{Gabriele Tocci}
\affiliation[Universit\"at Z\"urich]{\uzh}
\email{gabriele.tocci@chem.uzh.ch}
\title{Osmotic transport at the aqueous graphene and hBN interfaces: scaling laws from a unified, first principles description.
}

\abbreviations{}

\begin{document}

\begin{tocentry}
\includegraphics[width=9.0cm,height=4cm
,clip]{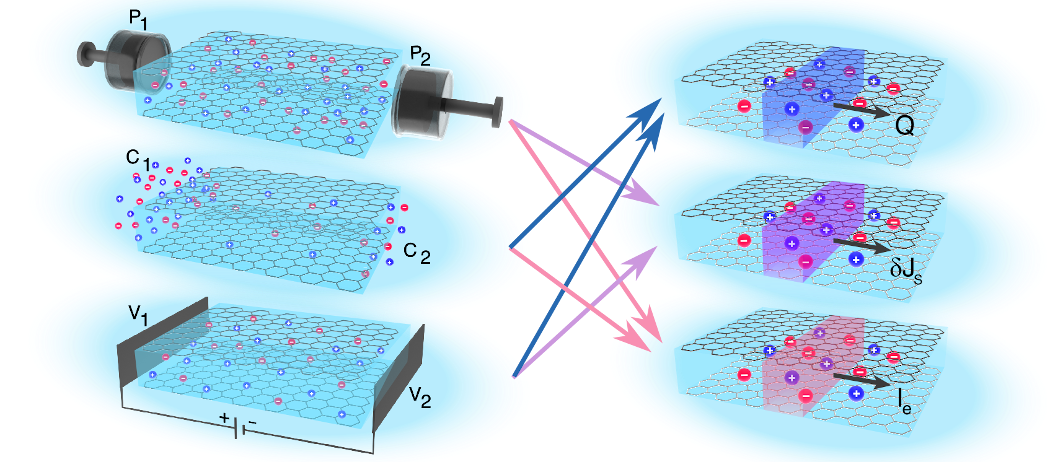}
\end{tocentry}

\begin{abstract}
Osmotic transport in nanoconfined aqueous electrolytes provides new venues for water desalination and ``blue energy'' harvesting; the osmotic response of nanofluidic systems is controlled by the interfacial structure of water and electrolyte solutions in the so-called electrical double layer (EDL), but
a molecular-level picture
of the EDL is to
a large extent still lacking.
Particularly, the role of the electronic
structure has not been considered
in the description of electrolyte/surface
interactions.
Here, we report enhanced sampling simulations
based on \textit{ab initio} molecular dynamics,
aiming at unravelling the free energy of prototypical
ions adsorbed at the aqueous graphene and hBN interfaces,
and  its consequences on nanofluidic osmotic transport.
Specifically, we predicted the zeta potential, the diffusio-osmotic
mobility and the diffusio-osmotic conductivity
for a wide range of salt
concentrations from the \textit{ab initio}
water and ion spatial distributions through an analytical
framework based on Stokes equation and a
modified Poisson-Boltzmann equation.
We observed concentration-dependent scaling laws,
together with
dramatic differences in osmotic transport
between the two interfaces, including
diffusio-osmotic flow and current
reversal on hBN,
but not on graphene.
We could rationalize the results for the three osmotic responses with a simple model based on characteristic length scales for ion and water adsorption at the surface, which are quite different on graphene and on hBN.
Our work provides first principles insights into the
structure and osmotic transport of aqueous electrolytes
on two-dimensional materials and
explores new pathways for efficient water
desalination and osmotic energy conversion.
\end{abstract}

\section{Introduction}
Universal access to drinkable water and
 widespread production of electricity
from renewable energy sources are two
of the most daring challenges faced by
modern society. Progress in the field of nanofluidics
offers alternative solutions to
water desalination and to energy conversion
through the mixing of salty and fresh water,
so-called ``blue'' energy harvesting.
Over the past decade, many novel osmotic
transport phenomena have been observed
in nanofluidic
systems.\cite{siria2017new,wang2017fundamental,Tong2021,macha20192d}
Among several noticeable examples is the osmotic
current generated across boron nitride nanotubes
and MoS$_2$ nanopores,
whose power densities exceed by several orders of magnitude
those produced by conventional
membranes\cite{Siria2013,feng2016single}.
A further example is the observation of qualitatively
different current-voltage characteristics
in the nonlinear transport of ions across graphene
and hBN angstrom-scale
slits, which interestingly hints at the critical role
of the crystal and electronic structure of the interface\cite{esfandiar2017size}.
The measurement of conductance oscillations and
Coulomb blockade in sub-nanometers MoS$_2$ pores
also indicates that  chemical nature,
 dimensions and geometry  of nanopores
are key factors to the observed nonlinear behaviours\cite{feng2016observation}.
Thus, it stands to reason that
obtaining a molecular-level picture of
aqueous interfaces is essential to predict and control
osmotic transport phenomena, and may lead to
fundamental advances in the field of nanofluidics.

A comprehensive picture of the structure of
water and electrolyte solutions
at electrified surfaces, in the
so-called electrical double layer (EDL),
has not been obtained so far.
Specifically, the structure of the EDL is not
accurately described using
the standard Gouy-Chapman theory
of the EDL based on the Poisson-Boltzmann (PB)
equation\cite{Hunter2001}.
For instance, in a thin
region of the EDL -- typically of the order of 1\,nm --
ions may interact specifically with solid
surfaces. Ion-specific
effects in this region, which we define ``adsorption layer'', are not captured
by the standard PB equation\cite{luo2006ion,huang2007ion}.
Additionally, the PB theory
of the EDL typically neglects the polar nature
of water and water layering at the liquid/solid
interface, and ignores
spatial and dynamic interface heterogeneities\cite{gonella2021water,Hartkamp2018,Markovich2016,Bonthuis2013,limmer2013hydration}.
Experimentally, a vast array of techniques have been used to probe
the structure of the EDL\cite{Hartkamp2018}. Recent examples include
second harmonic generation, which has
revealed structural and dynamical
heterogeneities in the water orientation
at the interface with silica\cite{macias2017optical},
and ambient pressure X-ray photoelectron spectroscopy,
which has probed the shape of the electrostatic potential profile of aqueous electrolytes
on gold electrodes at different concentrations\cite{favaro2016unravelling}.
Despite the tremendous advancement these types of work represent
for the field, achieving sub-nanometer resolution, which is required
to characterise the molecular structure of the EDL,
remains an open experimental challenge.

A further challenge is to link the  structure of aqueous  interfaces to osmotic transport properties. Although experiments have hinted at the predominant role of charged groups, pore geometry and pore chemistry\cite{Hartkamp2018}, a microscopic characterization of the interface under operating conditions has not been obtained.
Alternatively, atomistic simulations, and in particular molecular dynamics, can be used to explore the structure of the EDL at the molecular level.
Force-field molecular dynamics (FFMD), which is based on an empirical description of the interactions between the constituent atoms, has yielded invaluable insights into the structure of the EDL \cite{Siepmann1995InfluenceSystems, Scalfi2020ChargeEnsemble,scalfi2020a,Limmer2013ChargeCapacitors} and into the molecular mechanisms underlying liquid and solute transport in nanofluidics\cite{striolo2016carbon,phan2016confined,Faucher2019,Falk2010MolecularFriction,Ma2015WaterFriction,xie2018fast,huang2007ion,Ajdari2006,heiranian2015water,noh2020ion,liu2018pressure,simoncelli2018blue,Kalra2003OsmoticMembranes,Vasu2018ElectricallyMembranes}
Yet, it is challenging to determine accurate force fields that incorporate
electronic structure effects observed at complex liquid/solid interfaces.
In contrast,
\textit{ab initio} molecular dynamics (AIMD) simulations
based on density functional theory (DFT)
are instrumental to compare
the structure and dynamics of different aqueous
interfaces on equal footing.
Although AIMD is increasingly being used
to characterise the structure of
aqueous interfaces
\cite{le2020molecular,cheng2012alignment,gross2019modelling,lan2020ionization,seiler2018effect},
including OH$^{-}$ adsorption on two-dimensional
materials\cite{Grosjean2016},
H$_3$O$^+$ adsorption on TiO$_2$\cite{stecher2016first}
and ion adsorption at the liquid/vapour interface\cite{duignan2021toward,baer2011toward},
the impact of ion- and water-specific interactions
on osmotic transport at aqueous interfaces has not
been investigated with AIMD so far.

In this work we tackle two
challenges: we determine the structure of model aqueous
electrolyte interfaces from \textit{ab initio} methods and we compute the
mobilities underlying osmotic transport processes due to different applied external fields
(see Fig.~\ref{fig:schematic}). Our systems consist of a potassium iodide (KI) solution at the
aqueous graphene and  hBN interfaces.
We chose the aqueous graphene and hBN
interfaces because they are well-studied model systems in
nanofluidics and for their potential impact
as nano-osmotic power generators\cite{macha20192d}.
Also, we focus on KI
as a model electrolyte displaying ion-specific
effects\cite{schwierz2010reversed,huang2007ion,baer2011toward}.
The structure of the EDL and osmotic transport coefficients are computed
according to the following steps: First, we calculate the free
energy of adsorption
of K$^+$ and I$^-$ ions dissolved in a 2 nanometer-thick
water film using enhanced sampling
techniques based on AIMD  (a snapshot of a
representative system is shown in the inset of
Fig.~\ref{fig:free_energy}(a)); Second, we obtain the electrostatic potential profile and the
spatial distributions of the ions
at different salt concentrations by solving
a modified Poisson-Boltzmann (mPB) equation that
accounts for the ions' free energy of
adsorption on the sheets;
Finally, osmotic transport coefficients are obtained, within linear response theory,
by computing the relevant fluxes resulting from each external field, based on Stokes equation with a slip boundary condition at the wall.
We find remarkable ion- and surface-specific
adsorption of KI at the graphene and hBN interface. Such specific effects give rise
to concentration-dependent scaling laws of the
osmotic transport coefficients and result into strikingly
different osmotic transport behaviour at the graphene
and hBN interface. We rationalize the obtained scaling laws
with a theoretical model that describes
ion and water adsorption in terms of characteristic
length-scales that are limited to a few molecular diameters.

\section{Results}
\subsection{Unified framework of osmotic transport}
\begin{figure*}[thb!]
\includegraphics[width=\textwidth]{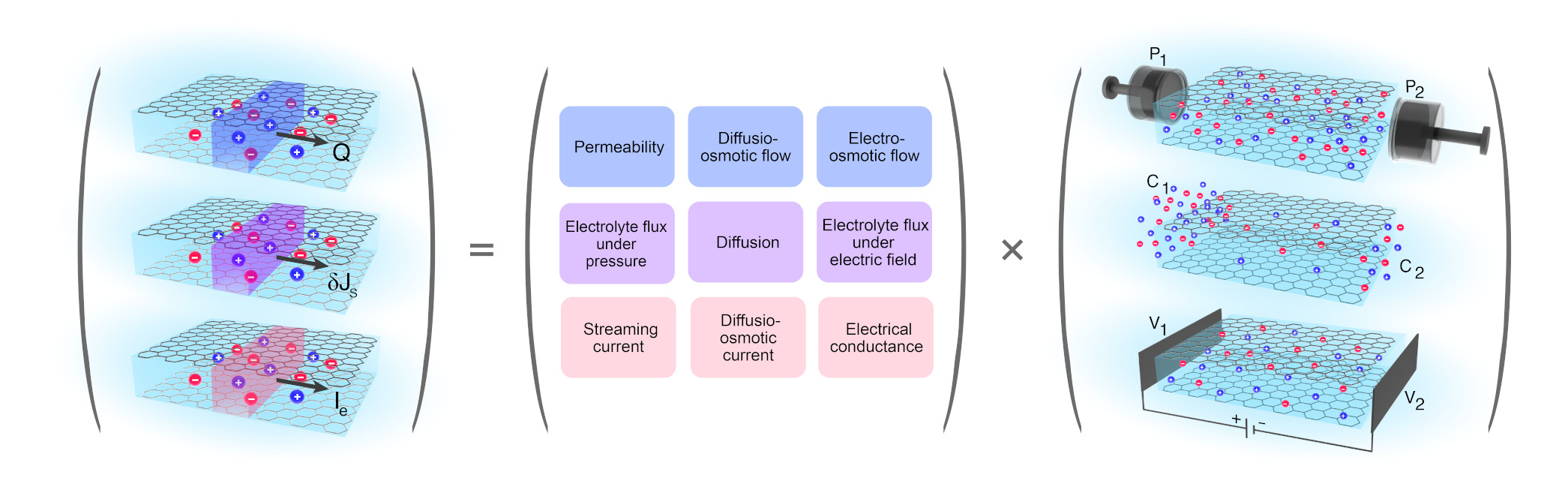}
\caption{\label{fig:schematic}
Schematic of the linear system of equations for osmotic transport.
The off-diagonal elements of Onsager
transport matrix represent the osmotic transport
coefficients, and are the central
quantities obtained in this work. They are computed
from the linear response of the fluxes,
(vector schematically shown on the left)
to an applied external force (vector on the right).
The matrix elements are color-coded according
to the color labeling of the respective flux, i.e., blue, violet and pink for
the elements due to a volumetric flow rate $Q$, an
excess solute flux $\delta J_s$ and an electrical current $I_e$, respectively.
From top to bottom, the external force vector on the right labels a pressure
gradient, a concentration gradient and an electrostatic potential gradient.
The sheets schematically depict a slit geometry, with cations and anions shown in
blue and red. Schematic inspired from Ref.~\citenum{marbach2019osmosis}. }
\end{figure*}

One of the main objectives of this work is to provide a theoretical
framework to calculate the osmotic transport coefficients
from the molecular structure of water and ions at liquid/solid interfaces.
We do so within linear response theory, where the
system of equations shown in Fig.~\ref{fig:schematic}
describes the thermodynamic fluxes resulting from applied external forces.
The desired transport coefficients
are the off-diagonal elements of the matrix
shown in the middle of Fig.~\ref{fig:schematic},
and obey Onsager's reciprocal relations\cite{Onsager1931a,Onsager1931b},
\textit{i.e.} the matrix is symmetric.
The osmotic transport coefficients
are computed from  hydrodynamics through
the Stokes equation, in which
the velocity profile is obtained
using a partial slip boundary condition at the wall $v(z=0) = b \partial _z v(z=0)$, with $b$
the slip length\cite{Bocquet2007}.
Throughout this work we thus assume
that continuum hydrodynamics
is valid at the nanoscale and
that the viscosity is homogeneous.
On hydrophobic, slipping surfaces such as the ones considered in this work, our assumptions have been shown to provide an accurate description of the velocity profiles even in the first molecular layers of the liquid.\cite{Bonthuis2013}

We start with electro-osmosis (EO), the flow generated by an electric field along a nanochannel, whose transport coefficient is illustrated in the top-right element of the matrix in Fig.~\ref{fig:schematic}.
The electro-osmotic response is commonly quantified by the so-called zeta potential $\zeta$, which relates the electro-osmotic velocity in the bulk liquid $v_\text{eo}$ to the electric field along the channel $E$ through the Helmholtz-Smoluchowski relation: $v_\text{eo} = -(\eps \zeta / \eta) E$, with $\eps$ and $\eta$ the permittivity and viscosity of the liquid in bulk, respectively.
According to Onsager's reciprocal relations, $\zeta$ also quantifies the streaming current density $j_e$ generated by a pressure gradient $-\nabla p$ along the channel: $j_e = -(\varepsilon \zeta / \eta) (-\nabla p)$.
From hydrodynamics equations,
one can relate the $\zeta$-potential to the charge density profile at the interface
(see \textit{e.g.} Refs.~\citenum{huang2007ion,marbach2019osmosis} and the supporting information (SI)):
\begin{equation}
\label{eq:zeta}
    \zeta = - \frac{1}{\eps} \int_0^{\infty}
    (z + b) \rhoe(z) \,\df z ,
\end{equation}
where the charge density
distribution is given by
$\rhoe = q_e\left(n_+ - n_-\right)$, with $n_+$ and
$n_-$ the cation and anion number
densities, respectively,
and $q_e$ the elementary charge.
Note that no assumption was made on the dielectric permittivity of the system: the bulk dielectric permittivity $\eps$ only appears in Eq.~\eqref{eq:zeta} through the Helmholtz-Smoluchowski definition of $\zeta$.

We move on to introduce diffusio-osmosis (DO), the flow generated by a gradient of salt concentration along the channel\cite{anderson1989colloid} (see Fig.~\ref{fig:schematic}). Diffusio-osmosis can be quantified by the so-called diffusio-osmotic mobility $D_\text{DO}$, which relates the diffusio-osmotic velocity in the bulk liquid $v_\text{do}$ to the gradient of salt concentration $\ns$: $v_\text{do} = D_\text{DO} (-\nabla \ns / \ns)$.
Employing again Onsager's reciprocal relations, $D_\text{DO}$ also quantifies the streaming excess solute flux density $\delta j_s$
generated by a pressure gradient
along the channel,
see details in Ref.~\citenum{Ajdari2006} and in the SI.

From hydrodynamics equations, one can relate $D_\text{DO}$ to the ionic density profiles $n_\pm(z)$ and to the water density profile $n_\text{w}(z)$ (normalized by its bulk value $n_\text{w}^\text{b}$) at the interface (see the SI):
\begin{multline}\label{eq:D_DO2}
    D_\text{DO} = \frac{\kt}{\eta} \int_0^{\infty} \left( z + b \right) \times \\ \left\{ n_+(z) + n_-(z) - 2 \ns \frac{n_\text{w}(z)}{n_\text{w}^\text{b}} \right\} \, \mathrm{d}z ,
\end{multline}
with $k_\text{B}T$ the thermal energy.
The integral expression for  $D_\mathrm{DO}$ in Eq.~\eqref{eq:D_DO2} is a first
central result of this work (see the derivation in the SI).
In particular, the contribution of the water density profile, which is usually ignored in the theoretical expressions of $D_\mathrm{DO}$ \cite{Mouterde2018,marbach2019osmosis}, makes a key difference, as ignoring it leads to a spurious negative contribution to the integral in the vacuum-like
region between the first water layer and the surface.
Note that an additional flow can be generated under a salt concentration gradient: for salt ions with an asymmetric diffusivity, a so-called diffusion electric field $E_0$ appears to avoid charge separation, creating an electro-osmotic flow, which adds to the intrinsic diffusio-osmotic flow\cite{anderson1989colloid,Lee2014b}. However, as discussed in the SI (see Fig.~S3), this
electro-osmotic component is negligible
in the systems considered here.

Finally, a gradient of salt concentration along the channel also generates an electric current, called diffusio-osmotic current (see Fig.~\ref{fig:schematic}), which is proportional to the perimeter of the channel cross section $P$. In order to quantify the intrinsic response of the liquid-solid interface, independently of the channel geometry, we therefore define the so-called diffusio-osmotic conductivity $K_\text{osm}$, which relates the diffusio-osmotic current generated per unit length of the channel circumference, $I_e/P$, to the gradient of salt concentration $\ns$: $I_e/P = K_\text{osm} (-\nabla \ns / \ns)$; note that slightly different definitions can be found in the literature\cite{Siria2013,marbach2019osmosis}.
According to Onsagers' reciprocal relations, $K_\text{osm}$ also quantifies the excess solute flux
generated by an electric field along the channel, see details in the SI.
From hydrodynamics equations, and assuming a homogeneous dielectric permittivity (this assumption will be discussed later in the article),
one can express $K_\text{osm}$ as:
\begin{multline}\label{eq:K_osm2}
K_\text{osm} = \frac{\kt q_e}{4 \pi \lB \eta} \int_0^{\infty}  \left[ \phi(z) - \phis -  \frac{2 \, \sgn(\Sigma) b}{\lGC} \right] \\ \times
\left\{ n_+(z) + n_-(z) - 2 \ns \frac{n_\text{w}(z)}{n_\text{w}^\text{b}} \right\} \,\mathrm{d}z,
\end{multline}
using the same notations as for Eq.~\eqref{eq:D_DO2}, and with $q_e$ the absolute ionic charge, $\phi(z) = q_e V(z) / (k_\text{B}T)$ the reduced electrostatic potential, $\phis$ its value at the surface, $\lB = q_e^2/(4\pi\eps\kt)$ the Bjerrum length (at which the
electrostatic interaction between two ions is comparable to the thermal energy),
and $\lGC = q_e/(2\pi \lB |\Sigma|)$ the Gouy-Chapman length\cite{Herrero2021}.
We remark that the expression for $K_\text{osm}$ shown in Eq.~\eqref{eq:K_osm2}
is the second important result of this work (see the SI for a complete
derivation).

The ions' density and the  electrostatic potential profiles
appearing in Eqs.~(\ref{eq:zeta}-\ref{eq:K_osm2})
are determined from the solution of the
Poisson-Boltzmann equation \cite{Hunter2001,Schoch2008},
which is used to describe the EDL near
electrified interfaces and is modified to include
the free energy of ion adsorption \cite{schwierz2010reversed,luo2006ion},
here computed from first principles simulations:
\begin{multline}\label{eq:mPB}
    \mathrm{d}^2_z \phi(z) = - 4 \pi \lB \left[ n_+(z) - n_-(z) \right] \\
    =- 4 \pi \lB  \ns \left[ e^{-\phi(z) - g_+ (z)} - e^{ \phi(z) - g_- (z) } \right],
\end{multline}
where the dimensionless free energies of ion adsorption
$g_{\pm}(z)= \Delta G_{\pm}(z) /(\kt)$ are the key terms that distinguish
Eq.~\eqref{eq:mPB} from the standard PB description of the EDL,
and which importantly enable the possibility for non-zero solutions of Eq.~\eqref{eq:mPB} even in the absence of charged surfaces,
as it is the case in our
aqueous graphene and hBN interfaces.
Although
the form of Eq. (4) assumes a constant value
of the dielectric permittivity $\varepsilon$, we have also
considered a step model of
the dielectric constant\cite{schwierz2010reversed,huang2007ion} and we have computed the transport coefficients within this model in the SI (see Fig.~S4).
Whereas the electro-osmotic
and diffusio-osmotic coefficients are
not affected by the particular choice
made for $\varepsilon$, the
magnitude of the diffusio-osmotic
conductivity is altered depending
on the  model used for the dielectric constant;
still, the scaling as a function of concentration  and the change of sign remain the same, thus not
affecting the conclusion of our work (see the SI).

\subsection{Ion adsorption and water density oscillations
from first principles}
\begin{figure}
\includegraphics[width=0.42\textwidth]{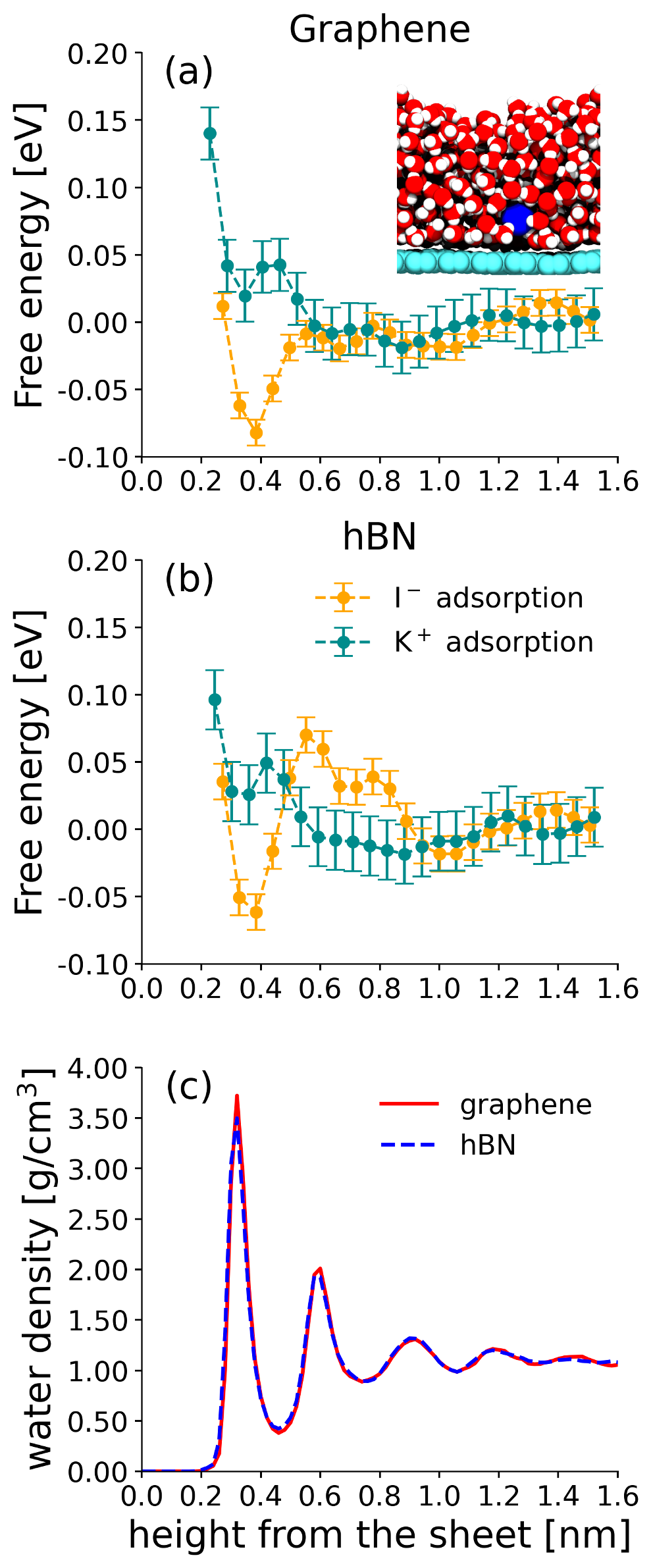}
\caption{\label{fig:free_energy}
Free energy of I$^-$ and K$^+$ adsorption
on (a) graphene and (b) hBN as a function of the height from the sheets,
and (c) water density profile on graphene and hBN.
The inset in (a) is a representative snapshot
of the water film simulation with a
iodide ion (blue) adsorbed on graphene (cyan).
The free energy profile of K$^+$
is similar on graphene and hBN and
does not show an adsorption minimum, in contrast with I$^-$ adsorption, which
reveals a minimum near the sheets
and a barrier on hBN, but not on graphene.}
\end{figure}

Thus, we start to discuss our simulation results by
presenting the free energy profiles
of I$^{-}$ and K$^{+}$ and the water density
profiles on graphene and hBN.
Figs.~\ref{fig:free_energy}(a) and (b)
display the free energy of ion adsorption
obtained from our \textit{ab initio} umbrella sampling
simulations. Significant ion- and surface-specific
adsorption can be observed,
which are limited to a
region of about 1\,nm from the sheets.
Further, the free energy of adsorption of the K$^+$
ion is essentially the same on
graphene and hBN, and whilst the signature of a local minimum appears
at a height of about 0.4\,nm from the sheets, K$^+$ is clearly more stable
in the bulk water region. On the other hand,
the free energy profile of I$^-$ exhibits pronounced differences
on graphene and hBN: a global minimum of about
$-0.08$ eV and $-0.06$\,eV is observed on graphene and hBN, respectively,
where I$^-$ is physisorbed at a height of about 0.4\,nm on both sheets.
Direct anion-substrate interactions due to van der Waals
dispersion forces are likely responsible for the
observed minimum, as reported in previous
FFMD simulations and second harmonic generation experiments
performed on closely related systems.\cite{mccaffrey2017mechanism}
Interestingly, an energy barrier of
about 0.07\,eV is observed for I$^-$
on hBN between the adsorption minimum at a height of 0.4\,nm
and the bulk water region at about 1\,nm from the sheet.
On graphene instead, a free energy barrier is not
observed. Analysis of the dipole orientation of water at the
two interfaces (reported in the SI in Fig.~S2) suggests that subtle
differences in the water dipole orientation
on the two sheets may be
responsible for the distinct features of the
free energy profile of I$^{-}$
on graphene and on hBN.
While some changes in the water dipole orientations on the
two sheets are noticed,
the water spatial density distribution
is remarkably similar on graphene and hBN and
presents strong oscillations
that are gradually suppressed above 1.0-1.2\,nm
from the surface, as highlighted in previous work\cite{tocci2014friction}.
In Eq.~\eqref{eq:mPB},
the ion density distributions, the
electrostatic potential profile $\phi(z)$
and its value at the sheets
$\phis$ depend non-linearly
on the bulk salt concentration $\ns$
through the free energy profiles shown in
Fig.~\ref{fig:free_energy}(a) and (b).
Both the free energy profiles of
the ions adsorbed on the sheets
and the water density profile $n_\text{w}(z)$
enter the transport integral expressions (Eqs.~(\ref{eq:zeta}-\ref{eq:K_osm2}))
and are therefore key to
understand osmotic transport at the interface.

\subsection{Concentration dependence of osmotic transport}
The observed ion- and surface-specific
adsorption and the layering  of the water density
profiles have pronounced effects
on the osmotic transport coefficients.
Slippage can also enhance osmotic
transport dramatically\cite{Ajdari2006}, as it can be
evinced from Eqs.~(\ref{eq:zeta}--\ref{eq:K_osm2}).
In this work, we used slip length values
taken from previous AIMD results\cite{tocci2020nanoscale}.
Water flows significantly faster
on graphene ($b=19.6$\,nm) than on hBN ($b=4.0$\,nm),
and we wish to explore the consequences this
bears for osmotic transport.

\begin{figure*}[thb!]
\includegraphics[width=1\textwidth]{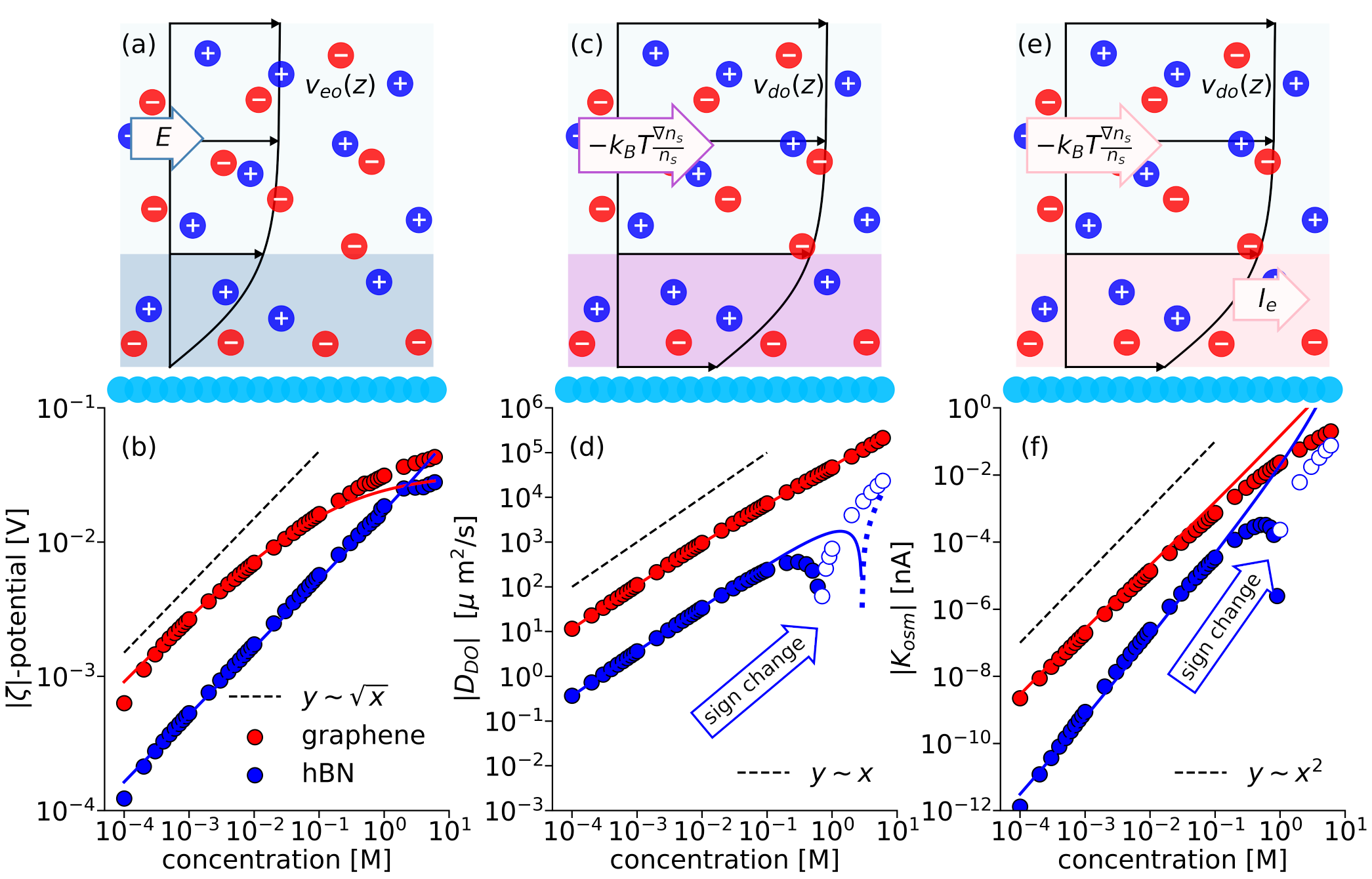}
\caption{\label{fig:transport_coefficients}
Molecular representation and concentration dependent
scaling of osmotic transport at the aqueous
graphene and hBN interfaces.
Schematic of the electro-osmotic velocity
profile $v_\mathrm{eo}(z)$ arising from
an electric field $E$ (a) and absolute
value of the $\zeta$ potential as a function of
salt concentration (b);
Schematic of the diffusio-osmotic velocity
profile $v_\mathrm{do}(z)$ in response to a concentration gradient $-k_\mathrm{B}T (\nabla\ns)/\ns$ (c) and absolute
value of the diffusio-osmotic coefficient $|D_\mathrm{DO}|$ (d);
Representation of the
diffusio-osmotic current $I_e$ arising from a concentration gradient
$-k_\mathrm{B}T (\nabla\ns)/\ns$ (e), and
absolute value of the diffusio-osmotic conductivity $|K_\mathrm{osm}|$ (f).
The transport coefficients display different
 scaling behaviours (see dashed lines).
The symbols are obtained from the numerical
integration of Eqs.~(\ref{eq:zeta}-\ref{eq:K_osm2}),
whereas the solid and dotted lines are from the effective surface charge model.
In (d) and (f) the arrows point to a
sign change, whereby $D_\text{DO}$
and $K_\text{osm}$
are negative on hBN at concentrations $\gtrsim 1$\,M, as indicated also
by the empty symbols and dotted line in (d).
In (a) $v_\mathrm{eo}(z)$ is not enhanced by slippage,
opposite to $v_\mathrm{do}(z)$ in (c) and (e), see text for details.}
\end{figure*}

The osmotic transport coefficients
are displayed in Fig.~\ref{fig:transport_coefficients}:
Each transport coefficient follows
a different asymptotic behaviour
as a function of salt concentration
and remarkable changes are noticed also
between graphene and hBN.
Fig.~\ref{fig:transport_coefficients}(a)
is a schematic representation of electro-osmosis
for the systems considered here, and Fig.~\ref{fig:transport_coefficients}(b)
shows the absolute value of the
$\zeta$-potential as a function of
concentration. The absolute value of the $\zeta$ potential
increases from about $0.6$\,mV to
$40$\,mV on graphene and
from about $0.1$\,mV to
$20$\,mV on hBN.
A non-zero value of the $\zeta$
potential even in absence of a surface
charge highlights the
role of ion-specific adsorption on electro-osmosis,
as reported in the past by
electro-phoretic experiments and FFMD
simulations\cite{huang2007ion,petrache2006salt}.
It is also worth noting that
the slip contribution in Eq.~\eqref{eq:zeta} writes $-b/\eps \int_0^\infty \mathrm{d}z \rhoe = b \Sigma / \eps$, so that it vanishes for neutral surfaces\cite{huang2007ion}.
Thus, based on the form of Eq.~\eqref{eq:zeta},
it can be evinced that
both the magnitude and the different concentration dependence
of $\zeta$ on graphene and hBN solely arise from
differences in the free
energy profile of adsorption
of I$^{-}$ on the two sheets,
given that the free energy of K$^{+}$
is very similar on graphene and hBN.

A schematic representation of diffusio-osmotic
flow is illustrated in
Fig.~\ref{fig:transport_coefficients}(c) and
the concentration dependence of the
diffusio-osmotic coefficient $D_{\text{DO}}$
is presented in Fig.~\ref{fig:transport_coefficients}(d).
Whereas $D_\mathrm{DO}$ scales linearly
on graphene at all concentrations, deviations from
a linear asymptotics is noticed
on hBN already around $10^{-1}$\,M, and  above
 $1$\,M $D_\mathrm{DO}$ becomes negative (a positive value of $D_\mathrm{DO}$ indicates that the
diffusio-osmotic flow proceeds from high to low
concentration and vice-versa for negative $D_\mathrm{DO}$). Noting also that the K$^{+}$
free energy profiles
and the water spatial distribution
are very similar on graphene and hBN,
it is clear that $D_\mathrm{DO}$
changes sign on hBN because of the
different free energy of adsorption of I$^{-}$
from graphene.
We note that a liquid flow that proceeds
towards a larger concentration
has been observed before for neutral solutes,
which only interact specifically with the surfaces.\cite{Lee2014b,Lee2017}.
Opposite to electro-osmotic transport,
diffusio-osmotic transport is amplified
by slippage even in the absence of a surface
charge (see Eq.~\ref{eq:D_DO2}) and
a strong slip-induced enhancement of
the diffusio-osmotic flow
has been reported before\cite{Ajdari2006}.
 The larger slip-length, along with the
  absence of an adsorption barrier
of I$^{-}$ are the two main reasons
why $|D_\mathrm{DO}|$ is larger on graphene than on hBN.

Finally, a schematic
of the diffusio-osmotic
current mechanism is presented in
Fig.~\ref{fig:transport_coefficients}(e),
and the  scaling behaviour of the diffusio-osmotic
conductivity is shown in Fig.~\ref{fig:transport_coefficients}(e).
Below $10^{-1}$\,M, the diffusio-osmotic conductivity
exhibits a different scaling behaviour
with salt concentration on graphene
and hBN, in contrast to what was observed in the
case of the diffusio-osmotic coefficient.
On graphene, $K_\mathrm{osm}$
scales more slowly than on hBN, but it
remains positive at all concentrations considered.
On hBN on the other hand, a maximum in $K_\mathrm{osm}$ is observed
just below 1\,M, above which an abrupt sign change is
observed, similar to what observed for $D_\mathrm{DO}$.
Finally, in Eq.~\eqref{eq:K_osm2},  we note that in contrast to
$D_\mathrm{DO}$, slippage does not contribute to $K_\mathrm{osm}$
in the absence of a surface charge, for which $1/\lGC \propto \Sigma = 0$.
This is because slip shifts the DO velocity profile by a constant amount $v_\text{slip}$, so that the corresponding electrical current writes: $I_e \propto v_\text{slip} \int_0^\infty \mathrm{d}z \rhoe \propto v_\text{slip} (-\Sigma)$, and vanishes for neutral surfaces.

\subsection{Scaling laws of the osmotic transport coefficients}
To rationalize the different scaling properties,
we introduce an effective surface charge (ESC) model
of the EDL. The ESC model is discussed in
full detail in the SI, but here we
present the essential idea.
In this model, the structure of the EDL
is mapped to a standard PB description,
where the effects associated to ion-specific
adsorption are treated as surface terms by introducing
an effective surface charge $\Sigma_\mathrm{eff}$.
$\Sigma_\text{eff}$ is defined as
$\Sigma_\text{eff} = q_e \ns \left (e^{-\phi_\mathrm{s}} K_+ - e^{\phi_\mathrm{s}}K_- \right)$ and it
depends on the bulk salt concentration $\ns$,
on the surface potential $\phi_\mathrm{s}$ and on the characteristic
lengths $K_{\pm}$, which quantify the excess or
depletion of ions near the sheets (see the definition in Table~\ref{table:1}).
It will be shown that the
density oscillations of interfacial
water also enter this modified formulation
and are relevant to capture
diffusio-osmotic transport.
The key to this simplified description is that
the region where ions and water interact specifically
with the sheets is very thin compared to the EDL.
As such, the ESC model is strictly
valid when there is a net separation between the
water and ion adsorption length-scales
and those of the EDL, given by the Debye length and
the effective Gouy-Chapman length, which we define here in terms
of the effective surface charge
$\lGC^\text{eff}=q_e/\left(2\pi\lB |\Sigma_\mathrm{eff}|\right)$.
In the following, we will apply the ESC model to
Eqs.~(\ref{eq:zeta}-\ref{eq:K_osm2})
to rationalize the scaling behaviours observed in
Fig.~\ref{fig:transport_coefficients}.

Starting from electro-osmosis, the $\zeta$-potential
can be expressed within the ESC model, in the limit of small reduced surface potentials $\phis \ll 1$ (see the SI), as:
\begin{equation}
\label{eq:zeta_mGC}
     \zeta \approx \frac{\kt}{q_e} \phis \approx \frac{\kt}{q_e} \cdot \frac{   K_+ - K_- }{  2\lambda_D + K_+ + K_- }.
\end{equation}
The results of this equation are shown as solid lines in
Fig.~\ref{fig:transport_coefficients}(b).
To understand the two different limits observed in
the figure, it is instrumental to inspect
the values of the ion-specific length-scales
$K_+$ and $K_-$ entering the equation, which are
listed in Table~\ref{table:1}.
For both graphene and hBN, $K_+$ is negative
and indicates a net depletion of cations near the sheets,
whereas $K_-$ is positive and indicates a net accumulation
of anions. Noting also that for hBN,
$|K_+ + K_-| \ll \lambda_\mathrm D$ at all
concentrations,  Eq.~\eqref{eq:zeta_mGC}
simplifies to $\zeta \approx \kt (K_+ - K_-)/(2\lambda_D q_e)$, such that
$\zeta$ scales as the inverse of the Debye length, or equivalently
as  the square-root of the salt concentration, \textit{i.e.}, $\zeta \sim 1/\lambda_\mathrm D \sim \sqrt{\ns}$.
For graphene on the other hand,
departure from $\zeta\sim\sqrt{\ns}$ is visible
already at a concentration above $10^{-3}$\,M
because the term $|K_++K_-|$ becomes comparable to $\lambda_\mathrm D$.
Therefore, approximately above
$10^{-3}$\,M the scaling of $\zeta$ on graphene is better
captured by the full Eq.~\eqref{eq:zeta_mGC}.
Beyond $10^{-1}$\,M for graphene and beyond 1\,M for hBN,
the prediction from Eq.~\eqref{eq:zeta_mGC} deviates
from the numerically integrated results
because the surface potential
becomes of the order of the thermal voltage
(\textit{i.e.} $\phis \gtrsim 1$).

\begin{table*}[t!]
  \centering
\begin{tabular}{ P{3.5cm}P{5.0cm}P{3.5cm}P{3.5cm}}
\hline
\hline
  \centering
 lengths [nm]&definition&graphene&hBN\Tstrut\Bstrut\\
 \hline
 \Tstrut\Bstrut
   $\bm{b}$&  &$\bm{+19.6}$&$\bm{+4.0}$\Tstrut\Bstrut\\
   $K_+$&$\int_0^{\infty}  [e^{-g_+(z)}-1]\mathrm{d}z$&$-0.229$&$-0.207$\Tstrut\Bstrut\\
   $\bm{K_-}$&$\int_0^{\infty}  [e^{-g_-(z)}-1]\mathrm{d}z$&$\bm{+1.974}$&$\bm{+0.177}$\Tstrut\Bstrut\\
   $K_w$&$\int_0^{\infty}  [ n_w(z)/n_w^{b} -1]\mathrm{d}z$&$-0.179$&$-0.165$\Tstrut\Bstrut\\
   $L_+$&$K_+^{-1}\int_0^{\infty} z[e^{-g_+(z)}-1] \mathrm{d}z$&$-0.363$&$-0.442$\Tstrut\Bstrut\\
      $\bm{L_-}$&$K_-^{-1}\int_0^{\infty} z[e^{-g_-(z)}-1] \mathrm{d}z$&$\bm{+0.459}$&$\bm{-0.099}$\Tstrut\Bstrut\\
   $L_w$&$K_w^{-1}\int_0^{\infty}  z[ n_w(z)/n_w^{b} -1]\mathrm{d}z$&$-0.031$&$-0.027$\Tstrut\Bstrut\\
\hline
\hline
\end{tabular}
\caption{Slip length $(b)$ and length-scales characteristic
of cation-specific (with subscripts ``$+$''),
anion-specific adsorption (with subscripts ``$-$'') and water
density oscillations (with subscript ``$w$'')
at the aqueous graphene and hBN interfaces,
along with their definitions.
The slip length and the anion length-scales are
in bold to highlight the stark differences between graphene and hBN.}
\label{table:1}
\end{table*}
Application of the ESC model to diffusio-osmosis
(Eq.~\eqref{eq:D_DO2}) yields the following
relation for $D_\text{DO}$ (derived in the SI):
\begin{multline}
\label{eq:D_DO_mGC}
    D_\text{DO}
    \approx \frac{\kt}{2\pi\lB\eta} \left\{ -\ln\left(1-\gamma^2\right) + \frac{b|\gamma|}{\lGC^\text{eff}} + \right. \\
    \left. \frac{1}{4\debye^2} \left[ e^{-\phis} K_+ L_+ + e^{\phis} K_- L_- -2 K_\text{w} L_\text{w} + \right. \right. \\  \left. b \left( e^{-\phis} K_+ + e^{\phis} K_- -2 K_\text{w} \right) \right] \bigg \}.
\end{multline}
The first two terms depending on $\gamma$, with
$\gamma = \tanh (\phis/4)$,
result from a calculation of
the diffusio-osmotic flow according to the standard PB description of
the EDL\cite{anderson1989colloid,Mouterde2018}
and arise from the region in the EDL beyond the adsorption layer, whereas
the remaining terms involving the ion length-scales
($K_\pm$ and $L_\pm$) and the water length-scales
($K_w$ and $L_w$) arise from specific
interactions within the adsorption layer.
Along with $K_{\pm}$, the additional ion-specific
length-scales $L_{\pm}$ contributing to Eq.~\eqref{eq:D_DO_mGC}
(see definition in Table~\ref{table:1})
represent the characteristic thicknesses over which cations and anions adsorb.
The water-specific length-scale $K_w$ is characteristic of
a net accumulation or depletion of water near the
sheets, while $L_w$ represents
the characteristic size
over which water accumulates/depletes
at the aqueous interface (see definition in Table~\ref{table:1}).
By analysing the relevant terms in
Eq.~\eqref{eq:D_DO_mGC}, we can
interpret the scaling behaviour observed in
Fig.~\ref{fig:transport_coefficients}(c),
while pointing out that in the equation
$\lambda_\text{D}$, $l^{\text{eff}}_{\text{GC}}$,
$\gamma$ and $\phis$ all depend on concentration.
First, the
terms depending on $\gamma$ are
negligible compared to the adsorption
layer terms at small surface
potentials ($\phis <1$)
and at concentrations
below $\sim 10^{-1}$\,M
because to leading order
$\gamma \sim \phis$ whereas $e^{\pm\phis} \sim 1$.
Therefore, at low concentrations
a linear dependence of $D_\text{DO}$
can be readily understood based on the dependence
of the adsorption term proportional
to $\lambda_\mathrm D^{-2}$
 by realizing that $D_\mathrm{DO} \sim \lambda_\mathrm D^{-2} \sim \ns$.
Also, examining the characteristic
length-scales
listed in Table~\ref{table:1}, one notices  that
the terms involving $e^{\mp \phis }K_{\pm} L_{\pm}$ and
$2 K_{\text{w}} L_{\text{w}}$  can be neglected
compared to the factor proportional to $b$
 because the slip length is much larger than
 $|L_\pm|$ and $|L_{w}|$.
As such, $D_\mathrm{DO}$ is enhanced by slippage through
the slip length $b$ and the magnitude of $D_\mathrm{DO}$
is larger on graphene than on hBN partly
because of the larger slip length
in the former material.
As the concentration is
increased around 1\,M the PB term and in
particular the slip contribution $b|\gamma|/\lGC^{\text{eff}}$
can no longer be neglected.
Interestingly, competing effects between
the $b|\gamma|/\lGC^{\text{eff}}$ term, the ions' adsorption
term and the molecular-scale oscillations
 of water can give rise to a sign reversal
 in $D_\text{DO}$ at a critical
 concentration.
Reversal of diffusio-osmotic flow is
indeed observed on hBN,
but not on graphene, due to the
differences in the slip length and in the
ion adsorption length-scales between the two materials
(see Table \ref{table:1}).
On graphene,
the ion-specific adsorption terms
proportional to  $e^{-\phis}K_+ + e^{\phis}K_+$ and
the term proportional to $b|\gamma|/\lGC^{\text{eff}}$
dominate over the water contribution proportional to
$2K_w$ at all concentrations
and the diffusio-osmotic
flow always proceeds in the direction from high to low
concentrations. On hBN, instead, above a
critical concentration of about 1\,M
the water term becomes the dominant contribution
and a flow reversal is observed.
A more detailed analysis on the scaling behaviour
of $D_{\text{DO}}$ is provided in Fig.~S5
in the SI, where the diffusio-osmotic coefficient has been explicitly decomposed into the standard
PB contribution and the adsorption layer contribution.

Finally, an approximate expression for the diffusio-osmotic conductivity
$K_\mathrm{osm}$ is obtained to understand the scaling behaviour observed
in Fig.~\ref{fig:transport_coefficients}(c) as:
\begin{multline}
\label{eq:kosm_mGC}
    K_\mathrm{osm} \approx - \frac{\kt \Sigma_\text{eff}}{2 \pi \lB \eta} \times  \bigg ( 1  -\frac{\text{asinh} (\chi)} {\chi} \\
  + \frac{ e^{-\phis} \tilde{K}_+\tilde{L}_+   + e^{\phis} \tilde{K}_-\tilde{L}_-   -2 \tilde{K}_\text{w} \tilde{L}_\text{w}  }{4 \lambda_\mathrm D^2} \bigg ).
\end{multline}
The term $1-\text{asinh} (\chi)/\chi$
arises from the region in the EDL
beyond the adsorption layer,
with $\chi=\debye/\lGC^\text{eff}$\cite{Siria2013,Mouterde2018}.
The adsorption layer also contributes to
$K_{\text{osm}}$ through the ion-specific length-scales $\tilde{K}_{\pm}$ and $\tilde{L}_{\pm}$, as well as $\tilde{K}_\text{w}$ and $\tilde{L}_\text{w}$.
Their interpretation is
analogous to the characteristic lengths
reported in Table \ref{table:1},
but we refer the reader to the SI (Table~S1) for
their definition and numerical values.

At concentrations between $10^{-4}$\,M and $10^{-1}$\,M,
 Eq.~\eqref{eq:kosm_mGC} reproduces
 the asymptotic
 behaviour of the diffusio-osmotic conductivity,
 which scales as $K_\mathrm{osm} \sim n_\mathrm s ^{p}$,
 where $p\approx 2$ for graphene and
 $p\approx 2.5$ for hBN.
 We discuss each term separately to explain
 the scaling behaviour observed in
 Fig.~\ref{fig:transport_coefficients}(f).
First of all, to leading order, the effective surface charge
scales linearly with the salt concentration
(\textit{i.e.}, $\Sigma_\mathrm{eff}\sim n_\mathrm{s}$). Secondly,
noting that $\lGC^\text{eff} \propto \Sigma_\mathrm{eff}^{-1}$,
the contribution to the diffusio-osmotic
conductivity arising from the region beyond
 the adsorption layer
 also scales linearly with salt concentration
 (\textit{i.e.}, $1-\text{asinh} (\chi)/\chi \sim n_\mathrm{s}$).
Thirdly, the term arising from the
 adsorption layer exhibits
 a first obvious linear dependence on the
 concentration through the Debye length
 since $\lambda_\mathrm{D}^{-2} \propto n_\mathrm{s}$. This is indeed
 observed for graphene,
 such that overall the diffusio-osmotic
 conductivity scales as $K_\mathrm{osm} \sim n_\mathrm s ^{2}$.
 A second more subtle dependence on salt concentration however, is observed on hBN,
 where the adsorption layer term in  Eq.~\eqref{eq:kosm_mGC} changes sign
around $5\times 10^{-3}$ M (see Fig.~S6 in the SI).
The contribution  to  the diffusio-osmotic conductivity coming from the adsorption layer
term and the $1-\text{asinh} (\chi)/\chi$
term can be either  suppressed or enhanced
depending on the sign of the former, and ultimately results into a scaling of
$K_\mathrm{osm} \sim \ns^{2.5}$.
Further, similarly to what was observed in the case of the diffusio-osmotic
flow in Fig.~\ref{fig:transport_coefficients}(d),
at concentrations beyond 1\,M a reversal in the
diffusio-osmotic current is  observed in the
numerical calculations for hBN, because the water
contribution dominates over
its ion counterpart, whereas on graphene a sign change is
not observed because the ion-specific adsorption
contribution dominates at
all considered concentrations.
We note that the assumption of a linear
electrostatic potential difference $\phi(z) -\phis$
made while deriving Eq.~\eqref{eq:kosm_mGC} breaks down
at large concentrations, thus Eq.~\eqref{eq:kosm_mGC}
fails to predict the sign change on hBN at around 1 M.
\section{Discussion}
In this section, we
discuss the significance of our
work in connection with theoretical and experimental results
on the structure of the EDL and of osmotic transport in nanofluidics.
In our approach to calculate the
spatial distribution of ions in the EDL,
the electrostatic potential and the
free energy profile of ion adsorption are
determined, respectively, from the
solution of the mPB equation and from our
enhanced sampling simulations.
This framework has been introduced in
the past to study the structure of the EDL
at liquid/liquid interfaces\cite{luo2006ion} and to
shed light on the Hofmeister
series at hydrophobic and hydrophilic
surfaces for different surface
charges with FFMD simulations\cite{schwierz2010reversed}.
However, deviations from a modified PB
description of the electrostatic
potential appear at concentrations of the order of 1 M\cite{gonella2021water},
and it would be interesting
to incorporate
more advanced theories into our framework
to ameliorate the deficiencies
underlying the mPB theory
at such concentrations\cite{duignan2021toward,hartel2015fundamental,netz2001electrostatistics,kardar1999friction}.

Despite their substantial computational cost,
our \textit{ab initio} simulations
have also revealed that ion adsorption
can be surface-specific, as illustrated
by the differences in the free
energy of I$^-$ on the two sheets but not
of K$^+$. In particular, the presence of a
free energy barrier of I$^-$ on
hBN and not on graphene can be ascribed to
 differences in the water dipole orientation
 on the two sheets.
Deep UV second harmonic generation,
already used to investigate the free energy
of adsorption of SCN$^-$ on
graphene\cite{mccaffrey2017mechanism},
is an ideal technique to explore
ion adsorption on other substrates, including hBN.
Also, X-ray photoelectron spectroscopy
would be instrumental to probe the shape of the electrostatic potential profile
at aqueous electrified interfaces\cite{favaro2016unravelling}.

A central contribution of this paper is to have provided
a unified framework of osmotic transport that can be computed
from the structure of water and ions at aqueous interfaces,
and which transport coefficients can be probed experimentally. Measurements of osmotic transport of KI
solutions across graphene or hBN have not appeared yet, but
we think that they are already possible,
given that measurements of ion transport across
{\AA}-size slits have been reported,
and that diffusio-osmosis of several types of salts across
silica surfaces has also been probed.
It would be desirable that such experiments
be performed at the point of zero charge,
since differences in pH lead
to variations in the surface charge\cite{secchi2016scaling,Grosjean2019,Siria2013,Mouterde2018} and thus would likely alter the scaling behaviour
of the transport coefficients with concentration.
For instance,
a different scaling behaviour of the electrical
conductivity on salt concentration
has been observed in carbon nanotubes. This has been captured by
charge regulation models\cite{secchi2016scaling, biesheuvel2016analysis}
which are, however,
phenomenologically different
from the mechanisms described here, as we remark that
the sheets are not charged.

In absence of experiments on KI solutions,
we can connect our work to recent streaming-voltage experiments of KCl solutions across graphitic
\AA-size slits\cite{mouterde2019molecular}. Such experiments
did not present a clear dependence of the $\zeta$-potential
as a function of concentration
in a range between $ 10^{-3}$\,M and $10^{-1}$\,M
and extracted a value for $\zeta$ at least 10 times larger than that
shown in Fig.\ref{fig:transport_coefficients} (a).
Possible explanations are that, although ion-specific effects
of a KCl solution are less pronounced than a KI solution,
under the extreme confinement regime probed
in such experiments the Stokes equation of
hydrodynamics breaks down. Additionally,
ion exclusion at the entrance of the
membranes might play an important role in \AA-scale slits.

Concerning diffusio-osmosis,
we discuss our results in connection with
measurements performed on silica surfaces,
where the diffusio-osmotic coefficient has
been measured for several types of aqueous electrolytes, including KI\cite{Lee2014b}.
The reported diffusio-osmotic coefficient of KI on silica
is $D_\mathrm{DO} \approx 250 \mu\text{m}^2/s$,
independent of concentration.
Instead, we observe approximately a linear dependence
at all concentrations for graphene, and
above $10^{-1}$\,M $D_\mathrm{DO}$ reaches values beyond
$10^{4}\,\mu\text{m}^2/\text{s}$.
The observed enhancement of
the diffusio-osmotic flow
of KI on graphene compared to
silica can be largely attributed to the larger slip length of graphene.
Although the \textit{ab initio} values
for the slip length on graphene and
hBN\cite{tocci2020nanoscale}
compare well to recent flow  experiments performed on graphene slits\cite{xie2018fast},
as well as on hBN nanotubes and on carbon nanotubes
with a large radius $R$ of 50\,nm\cite{secchi2016massive},
the slip length for carbon nanotubes tubes
with a smaller radius $R=20$\,nm
is about one order of magnitude
larger than that of graphene, $b\sim 200$\,nm.
Provided that the structure of the EDL
would be only modestly affected
by such a large nanotube radius, one can extrapolate
the value of the diffusio-osmotic
coefficient that would be obtained for carbon
nanotubes with $R=20$\,nm from our results on graphene.
Doing so would result in approximately a $10$-fold
enhancement in the diffusio-osmotic
coefficient because of the much larger slip length.
Additionally, compared to silica,
where the diffusio-osmotic flow proceeds
from high to low concentrations,
we observe a flow reversal on hBN.

Measurements of the diffusio-osmotic current of KI solutions on graphene or hBN
have also not been reported. As such, we
connect our results for $K_\mathrm{osm}$
to osmotic energy conversion experiments of KCl solutions
flowing across single-pore BN
nanotubes\cite{Siria2013}.
On BN nanotubes, the diffusio-osmotic conductivity
$K^\mathrm{BNNT}_\mathrm{osm}$ has
been extracted from the ratio between the
diffusio-osmotic current and the difference in
the salt concentration at the reservoirs as
$K^\mathrm{BNNT}_\mathrm{osm} = I_\text{osm} /(\Delta \ns /\ns)$.
In order to compare directly to our results, we
normalize the current with respect to the perimeter
$P= 2\pi R$ and the length $L$ of the nanotube, \textit{i.e.}
$K^{\prime \mathrm{BNNT}}_\mathrm{osm} = \frac{I_\text{osm} /P}{\Delta \ns /( \ns L)}$. With an experimental value of the nanotube length   $L=1250$\,nm and of the radius $R=40$\,nm, the equivalent
experimental value for the diffusio-osmotic conductivity is
$K^{\prime \mathrm{BNNT}}_\mathrm{osm} = K^\mathrm{BNNT}_\mathrm{osm} L/(2\pi R) = 0.35 - 0.80$\,nA. These values for the diffusio-osmotic conductivity are much
larger than those computed here (at any salt concentration)
because of the very large surface charge $\Sigma$
that was reported in experiments ($\Sigma \approx 0.1 -1$\,C/m$^2$).
In BN nanotubes, the diffusio-osmotic
conductivity has been found to scale linearly on
the pH of the solution, and thus on the surface charge,
but to be independent of salt concentration\cite{Siria2013}.
Here instead, the diffusio-osmotic conductivity
scales roughly as $K_\mathrm{osm} \sim \ns^{p}$
with an exponent $p\approx 2$ and $p\approx2.5$
for graphene and hBN, respectively, over several decades
of salt concentration. The sign reversal in the
diffusio-osmotic current, as well as
in the diffusio-osmotic flux,
could be particularly relevant in biology.
Similar mechanisms might be at work in
membrane proteins, which could exploit changes
in concentration to regulate charge and solute
transport\cite{van2006claudins,van2003reversal}.
Although the focus here has been devoted to the understanding of
osmotic transport across a single-pore membrane or channel,
further challenges lie ahead of diffusio-osmosis in order to
establish itself as a viable source of renewable energy, in
particular, for what concerns the generation of electricity using
multi-pore systems.\cite{macha20192d,Tong2021,Wang2021}

In conclusion, we have provided a unified description of
osmotic transport by coupling first principles simulations with a
mean field description of the EDL and with the Stokes equation
of hydrodynamics, and have applied this framework to understand
the osmotic transport properties of a
prototypical salt that displays pronounced
ion-specific effects on two-dimensional materials.
Through the mPB equation of the EDL (see Eq.~\eqref{eq:mPB}),
the transport coefficients in
Eqs.~(\ref{eq:zeta}-\ref{eq:K_osm2}) can
be readily computed from the spatial
distribution of ions and water
at the interface.
We have reported on a concentration-dependent
scaling behaviour of the osmotic transport coefficients
at the aqueous graphene and hBN interfaces
and explained it with a model that
accounts for ion-specific adsorption and
water layering in a thin region of the EDL
 of the order of 1\,nm.
The observed scaling, along with the possibility of
diffusio-osmotic flow and
current reversal, may provide additional routes
to further improve osmotic energy conversion.
Moreover, it could foster the development of nanofluidic
diodes and sensors and may be instrumental to shed light
on the mechanisms underlying
charge and solute transport across membrane proteins.

\section*{Materials and Methods}
\subsection*{Electronic structure and \textit{ab initio} molecular dynamics}
The \textit{ab initio} umbrella sampling
simulations are performed with
the CP2K code \cite{kuhne2020cp2k}, and the
electronic structure problem is solved using DFT
with the optB88-vdW functional \cite{jiri_solids,jiri_molecules}.
The optB88-vdW functional has been applied to investigate slippage
on two-dimensional materials \cite{Joly2016,tocci2020nanoscale}
and it describes the  structure of graphite and bulk hBN
accurately \cite{graziano_vdw_gra_bn}.
Despite the limitations of most density functionals
in describing water, the optB88-vdW functional appears
to be one of the most satisfactory \cite{gillan2016perspective}.
Reference quantum monte carlo calculations
of water monomers adsorbed on graphene and hBN sheets
\cite{brandenburg2019physisorption,al2015communication}
also show that this functional captures the relative stability
of water on different adsorption sites.

The free energy of adsorption of K$^+$ and I$^-$ ions
and the spatial water distribution were calculated from umbrella sampling simulations\cite{torrie1977nonphysical}.
The systems consist of water films containing 400 molecules and approximately 2 nm-thick
placed above a $2.56 \times 2.46$ nm$^2$ graphene and a $2.61 \times 2.51$ hBN nm$^2$ sheet.
Separate umbrella sampling simulations were performed for potassium and iodide
adsorption by restraining each ion at different heights  $z_0$ above the sheets with
a harmonic bias potential $U_b(z,t) = k_b/2 (z(t)-z_0)^2$,  $z(t)$ being the
instantaneous height of the ion above the sheets,
and $k_b = 836.8$ kJ/mol/nm$^2$ the \textit{spring} constant.
A total of 23 umbrella sampling windows were used for each system
and for each window the dynamics were propagated for 40 ps
in the NVT ensemble at 300\,K within the Born-Oppenheimer
approximation, except for I$^{-}$ adsorption on graphene,
for which  dynamics were propagated for additional 30 ps to test
a possible dependence of our results on the length of the
simulations. The free energy was reconstructed using umbrella integration\cite{kastner2011umbrella}.
Further computational details on the  optimization
of the wave-function and on the umbrella sampling
simulations are reported in the SI.
The tools used to perform the analysis and the CP2K input files
to reproduce the main results of the manuscript have been deposited on
GitHub and are available at \url{https://github.com/gabriele16/osmotic_transport_scaling_laws}.

 \begin{acknowledgement}
 GT is supported by the SNSF project PZ00P2\_179964.
 LJ is supported by 
the Institut Universitaire de France.
RM and GT acknowledge funding by the Deutsche Forschungsgemeinschaft (DFG, German Research Foundation) -- 390794421.
 We also thank the Swiss National Supercomputer Centre (CSCS) under PRACE for awarding us access to
 Piz Daint, Switzerland, through projects pr66 and s826.

 \end{acknowledgement}

\begin{suppinfo}
Further computational details and tests
on the structure and dynamics
of water at the interface with
graphene and hBN  from
force field and \textit{ab initio} simulations;
derivation of integral expressions for the transport coefficients;
details of the modified Poisson-Boltzmann description;
details of the effective surface charge model.
\end{suppinfo}

\providecommand{\latin}[1]{#1}
\makeatletter
\providecommand{\doi}
  {\begingroup\let\do\@makeother\dospecials
  \catcode`\{=1 \catcode`\}=2 \doi@aux}
\providecommand{\doi@aux}[1]{\endgroup\texttt{#1}}
\makeatother
\providecommand*\mcitethebibliography{\thebibliography}
\csname @ifundefined\endcsname{endmcitethebibliography}
  {\let\endmcitethebibliography\endthebibliography}{}

\end{document}